\documentclass{ws-procs961x669}            % default, citations in superscript

\newcommand{\be}{\begin{equation}}
\newcommand{\ee}{\end{equation}}
\newcommand{\bea}{\begin{eqnarray}}
\newcommand{\eea}{\end{eqnarray}}

\begin{document}
\title{New partial resummation of the QED effective action}

\author{Silvia Pla$^*$ and Jos\'e Navarro-Salas$^\dagger$}% and Silvia Pla$^*$}

\address{Departamento de Fisica Teorica and IFIC, Centro Mixto Universidad de Valencia-CSIC.\\
Facultad de Fisica, Universidad de Valencia, Burjassot-46100, Valencia, Spain.\\
$^*$E-mail: silvia.pla@uv.es
$^\dagger$E-mail: jnavarro@ific.uv.es}
%\author{A. N. Author}

%\address{Group, Laboratory, Street,\\City, State ZIP/Zone, Country\\E-mail: an\_author@laboratory.com}

\begin{abstract}
%The goal of this talk is to 
We explain a conjecture which states that the proper-time series expansion of the one-loop effective Lagrangian of quantum electrodynamics 
can be  partially summed in all terms containing  the field-strength invariants $\mathcal{F} = \frac{1}{4} F_{\mu\nu}F^{\mu\nu} (x)$, $\mathcal{G}= \frac{1}{4} \tilde F_{\mu\nu}F^{\mu\nu}(x)$. %, including those that also have  derivatives of the electromagnetic field strength. . 
This summation is encapsulated in a factor with the same form as the (spacetime-dependent) Heisenberg-Euler Lagrangian density. We  also discuss some implications and a possible extension in presence of gravity. We will focus on the scalar field case. %This talk is based on the article\cite{NP21}: Phys.Rev.{\bf D} 103 (2021) 8, L081702.\\

\end{abstract}

\keywords{One-loop effective Lagrangian; Partial resummation; Heisenberg-Euler Lagrangian; asymptotic expansion;}

\bodymatter
\section{Introduction}
%This proceeding is based on a talk given in the Sixteenth Marcel Grossmann Meeting \cite{talk}. 
In this contribution{\footnote{Expanded version of the talk given by S.P. in the Sixteenth Marcel Grossmann Meeting (2021).}}, %talk, 
we give more details on a conjecture proposed in Ref. \cite{NP21}, that claims that it is possible to make a factorization in the proper-time series of the one-loop effective Lagrangian of quantum electrodynamics (QED) that captures all explicit dependence on the field-strength invariants $\mathcal{F}$ and $\mathcal{G}$. %Here, we are going to explain in detail the content of the talk. 
The text is organized as follows. First, in Sections \ref{sect-EH} and \ref{gravity} we review some basics about the one-loop effective Lagrangian and its proper-time series expansion. We focus on the scalar QED effective Lagrangian and also on the corresponding effective Lagrangian in presence of gravity. We emphasize the existence of a modified, asymptotic expansion (also called $R$-summed expansion) in the gravitational scenario. In Section \ref{conjecture} we explain our conjecture for scalar QED (a sketch for spinor QED is given in Ref. \cite{NP21}). We also check it up to order $n=6$ (the last available coefficient in the literature). In Section \ref{consequences} we explain some physical consequences that can be easily obtained using our results. Finally, in Section 4 we give a brief summary of this contribution.  For simplicity, we will focus on the scalar QED case. In Appendices \ref{apA} and \ref{apB} we give the coefficients of the usual and the resummed proper-time series expansions for scalar QED.

\subsection{Heisenberg-Euler Lagrangian and the proper-time series expansion} \label{sect-EH}
The Heisenberg-Euler Lagrangian  captures the one-loop quantum corrections to the classical electromagnetic Lagrangian for quantum electrodynamics%(QED)
. It is obtained by integrating out the matter degrees of freedom, keeping the strength of the electromagnetic background constant \cite{Dunne1,Schwartz}. The intrinsic nonlinearities of the quantum corrections have important implications, such as pair creation from vacuum and vacuum polarization \cite{Schwinger}, or light by light scattering or vacuum polarization (see, for example, Ref. \cite{dunne3} and references therein). It was originally derived by Heisenberg and Euler \cite{HE} for spinor QED in 1936. A few months later, it was obtained by Weisskopf \cite{Weisskopf} for scalar QED  (see Ref. \cite{Vanyashin-Terentev} for a charged spin 1 field). These two theories are regarded as the paradigm of the modern effective theories in Quantum Field Theory. From now on, we will focus on the scalar QED Heisenberg-Euler Lagrangian. It is usually written in terms of the proper time parameter $s$, but, apart from this, it be written in several ways. By convenience, here we use the following form borrowed from Ref. \cite{parker-toms}: 
\be  \mathcal {L}_{scalar}^{(1)}=-\frac{1}{ (4\pi)^{2}} \int_0^\infty \frac{ds}{s^3} e^{-im^2 s}  \Bigg[\det \left( \frac{esF}{\sinh (esF)}\right )\Bigg]^{1/2}  , \label{Lscalar}  \ee
where $F\equiv F^\mu_{\,\,\, \nu}$. As we have stressed before, this result was obtained for constant electromagnetic backgrounds, but for arbitrary background configurations the form of the effective action is, in general,  {\it unknown}.

However, it is possible to build a general, asymptotic expansion for the one-loop effective Lagrangian in terms of the proper-time parameter. Here we use the expansion given in Ref. \cite{coefs1}, obtained via the string-inspired method in the world-line formalism (see Refs. \cite{coefs1, coefs2, coefs3,Schubert1} and also Refs.\cite{string-inspired1,string-inspired2}), namely
		\bea  \label{gs}\mathcal {L}_{scalar}^{(1)}&=& \int_0^\infty \frac{ds}{s} e^{-im^2 s} g(x;is) \ , \eea
with 
\begin{equation} \label{QED-expansion}
g(x ; i s)=\frac{1}{(4 \pi i s)^{2}} \sum_{n=0}^{\infty} \frac{(-i s)^{n}}{n !} O_{n}(x) \longrightarrow\left\{\begin{array}{l}
O_{0}=1, \quad O_{1}=0, \\
O_{2}=-\frac{ e^{2}}{6} F_{\kappa \lambda} F^{\kappa \lambda} \\
O_{3}=-\frac{ e^{2}}{20} \partial_{\mu} F_{\kappa \lambda} \partial^{\mu} F^{\kappa \lambda} \vspace{0.2cm}\\
\hspace{1.3cm} {\tiny (\ \cdots )}
\end{array}\right\}
\end{equation}
The expression above consist of an expansion in the number of external fields and the number of derivatives, and captures some general behaviour of the unknown formal one-loop effective Lagrangian for arbitrary backgrounds. The coefficients $O_n$ are  gauge-invariant and have mass dimension $2n$. They have been obtained up to 12th adiabatic order ($n=6$), and their length grows substantially with $n$. In Appendix \ref{apA} can be found the complete list of coefficients. We would like to stress that this expansion presents some clear advantages with respect to other proposals. In particular, it is written in the most compact form possible. Using the Bianchi identity, the antisymmetry of $F_{\mu \nu}$ and also integration by parts, the gauge-invariant coefficients $O_n$ have been simplified to the so-called minimal basis \cite{Muller}.

\subsection{One loop effective Lagrangian in presence of gravity} \label{gravity}

The role of the classical electromagnetic background can be replaced by a gravitational field, which is naturally coupled to quantized matter fields. In other words, it is possible to obtain quantum corrections to the Einstein-Hilbert Lagrangian induced by a quantum scalar field.  In general, the form of the effective action is not known, but it is also possible to make an asymptotic expansion of it \cite{DeWittbook}, as in the previous case.

In this context, the expansion of the one-loop effective Lagrangian is usually obtained in terms of the Schwinger proper-time expansion of the Feynman propagator \cite{Schwinger}, closely connected to the heat-kernel expansion and related techniques \cite{Vassilevich}. Specifically, the one loop effective Lagrangian in terms of the heat-kernel reads
\be \mathcal {L}^{(1)}_{scalar} =    \int_0^\infty \frac{ds}{s} e^{-im^2 s}  \langle x, s| x, 0 \rangle \ . \label{Skernel} \ee
In this expression, $\langle x', s| x, 0 \rangle$ is the kernel or ``transition amplitude" associated to the Feynman propagator, which obeys a Schrodinger-type equation in terms of the proper time parameter with appropriated boundary conditions, and  $\langle x, s| x, 0 \rangle$ represents the coincident limit $x'\to x$. In general, the kernel admits a general asymptotic expansion in terms of the proper time parameter, i.e.,
	 \begin{equation} \label{expansion - 1}
\langle x, s| x, 0 \rangle=\frac{1}{(4 \pi i s)^{2}} \sum_{n=0}^{\infty}(i s)^{n} a_{n}(x). %\longrightarrow {\scriptsize \begin{aligned}
%&\text { {\bf DeWitt coefficients.} They are local, } \\
%&\text { covariant, and gauge-invariant } \\
%&\text { quantities of mass dimension 2n }
%\end{aligned}} \nonumber
\end{equation}
The coefficients $a_n(x)$ are called DeWitt coefficients. These quantities of mass dimension 2$n$  are local, covariant and gauge invariant \cite{DeWittbook,gilkey}. We recall here that for QED, the expansion $g(x;is)$ [see Eq. \eqref{expansion - 1}] coincides with the heat kernel expansion, up to total derivatives 
\be g(x;is) = \langle x, s| x, 0 \rangle + total \ derivatives. \ee

Although the general form of the gravitational one-loop effective Lagrangian is not known, there is one special case where it can be evaluated exactly \cite{Bekenstein-Parker}: the Static Einstein Universe. In this case, the kernel $ \langle x, s| x, 0 \rangle$ reads
   \be
   \langle x, s| x, 0 \rangle = \frac{e^{-i(\xi-\tfrac{1}{6}) R s}}{(4\pi i s)^2}. \ee
where $R$ is the Ricci scalar and $\xi$ is the coupling constant to the scalar field. This is a non-perturbative result, that involves all powers of the proper time parameter $s$. From this result, Parker and Toms proposed in 1985 a new asymptotic expansion for the kernel \cite{Parker-Toms85}, defined as 
    \bea \label{R-summed exp}
  \langle x, s| x, 0 \rangle =\frac{ e^{-i(\xi-\frac{1}{6}) R(x) s} }{(4 \pi i s)^{2}} \sum_{n=0}^{\infty}(i s)^{n} \bar a_{n}(x).
   \eea     
		    The new adiabatic expansion in terms of the coefficients $\bar a_n(x)$ has an important advantage with respect to the previous one: it does not contain any term that vanishes when $R(x)$ is replaced by zero (see Ref.\cite{Jack-Parker} for detailed proof). In other words, the non-perturbative exponential factor captures the  exact dependence on the Ricci scalar in the generic Schwinger-DeWitt asymptotic expansion.	This factorization has some important physical consequences, related, for example, to the effective dynamics of the Universe\cite{set1} or to the curvature dependence in the running of the gauge coupling constants\cite{CJP}.

	The key idea is that, from a solvable case, it is possible to extract a non-perturbative factor in the general asymptotic expansion of the one-loop effective Lagrangian that completely captures
		  its dependence with the curvature scalar $R(x)$.  In this context, a natural question arises: is it possible to find a similar factorization for the QED Lagrangian? %The answer to this question leads to the conjecture that we propose in the next section.
\section{The conjecture} \label{conjecture}

  Based on the gravitational case, we propose the following  conjecture for the electromagnetic case \cite{NP21}: 
	{\it  The proper-time asymptotic expansion of the QED effective Lagrangian admits an exact resummation in all terms involving the field-strength invariants $\mathcal{F}=\tfrac{1}{4}F_{\mu \nu}F^{\mu \nu}$ and $\mathcal{G}=\tfrac{1}{4}\tilde F_{\mu \nu} F^{\mu \nu}$. The form of the factor is just the Heisenberg-Euler Lagrangian for QED, where the electric and magnetic fields depend arbitrarily on space-time coordinates}, namely

		 	\be  \mathcal {L}_{scalar}^{(1)}= \int_0^\infty \frac{ds}{s} e^{-im^2 s} {\small \Bigg[\det \left( \frac{esF(x)}{\sinh (esF(x))}\right )\Bigg]^{1/2}}  \bar g(x;is)    \ee
		 	[to compare with \eqref{expansion - 1}], where  the new asymptotic expansion $\bar g(x;is)$ reads
		      \be \label{expansion - 2}
		      \bar g(x ; i s)=\frac{1}{(4 \pi i s)^{2}} \sum_{n=0}^{\infty} \frac{(-i s)^{n}}{n !} \bar O_{n}(x), 
		      \ee
and where the coefficients $O_n(x)$ do not have terms that vanish when the electromagnetic invariants $\mathcal{F}(x)$ and $\mathcal{G}(x)$ are replaced by zero.
The asymptotic expansions $g(x ; i s)$ and $\bar g(x ; i s)$ are related by
		      \be \label{relation - 1}
		      g(x;is)={\small \Bigg[\det \left( \frac{esF(x)}{\sinh (esF(x))}\right )\Bigg]^{1/2}}  \bar{g}(x;is).
		      \ee
		      In summary, we have proposed a new adiabatic expansion $\bar g(x;is)$ that can be built from the previous one via Eq. \eqref{relation - 1}, and we have conjectured that, if the pre-factor is just the Euler Heisenberg Lagrangian, the new coefficients $\bar O_n$ associated with the new expansion will not have terms that vanish when the electromagnetic invariants are replaced by zero.
% In other words, if the conjecture works, the coefficients $\bar O_n$ will not depend on the electromagnetic invariants. 

In the last part of the section, we are going to compute the new coefficients and check that, effectively, our conjecture is satisfied. %We are going to devote the last part of the section to compute them and to check that, effectively, our conjecture is satisfied. 
First, we should expand the Heisenberg-Euler determinant in terms of the proper time,
\be
{\small \Bigg[\det \left( \frac{esF(x)}{\sinh (esF(x))}\right )\Bigg]^{1/2}}\sim 1+ U_2(x)(-i s)^2+ U_4(x)(-i s)^4+ U_6(x)(-i s)^6+\cdots , \label {det-expansion}
\ee
where 
 \bea
U_2(x)&=&\tfrac{e^2}{12}\operatorname{Tr}(F^2),\\
U_4(x)&=&\tfrac{e^4}{288}\operatorname{Tr}(F^2)^2+\tfrac{e^4}{360}\operatorname{Tr}(F^4),\\
U_6(x)&=&\tfrac{e^6}{10368}\operatorname{Tr}(F^2)^3+\tfrac{e^6 }{4320}\operatorname{Tr}(F^2) \operatorname{Tr}(F^4)+\tfrac{e^6}{5670}\operatorname{Tr}(F^6).
\eea
we note here that the trace $\operatorname{Tr}(F^{2n})$ can be always written in terms of powers of the electromagnetic invariants $\mathcal{F}$ and $\mathcal{G}$. For example, $\operatorname{Tr}(F^{2})=-F_{\mu \nu}F^{\mu \nu} = - 4 \mathcal{F}$. Now, inserting expansions \eqref{det-expansion}, \eqref{QED-expansion} and \eqref{expansion - 2} in Eq. \eqref{relation - 1} we find
  \be
	 O_0+ O_1+ (-i s)+\frac{O_2}{2}(-is)^2... = (1 +U_2 (-is)^2+...)\cdot ( \bar O_0+  \bar O_1 (-i s)+\frac{ \bar O_2}{2}(-is)^2+ ...)
	 \ee
	 and equating order by order we arrive to $\bar O_0=1$, $\bar O_1=0$,
%	 \bea	  O_2 &=& \bar O_2 + 2 U_2\,; \,\,\, \bar O_2=0, \\	  O_3 &=& \bar O_3,\\	  O_4 &=& \bar O_4 + 4! \, U_4,\\	  O_5 &=& \bar O_4 + 20\, U_2 \bar O_3,\\	  O_6 &=& \bar O_6 + 6! \, U_6 +  30\, U_2 \bar O_4.	 \eea
	 \bea
	 \bar O_2 &=&  O_2 - 2 U_2=0, \\
	 \bar O_3 &=&  O_3,\\
	 \bar O_4 &=&  O_4 - 4! \, U_4,\\
	 \bar O_5 &=&  O_4 - 20\, U_2 O_3,\\
	 \bar O_6 &=&  O_6 - 6! \, U_6 -  30\, U_2 \bar O_4.
	 \eea 
Finally, substituting the values of the coefficients $U_n$ and $O_n$	in the previous equations we arrive at the explicit values of $\bar O_n$, which can be found in Appendix \ref{apB}. It is easy to see that our conjecture is satisfied. Comparing the coefficients $O_n$ (Appendix \ref{apA}) with $\bar O_n$, one sees that all terms going with $\operatorname{Tr}(F^{2n})$ [or, equivalently, with $\mathcal{F}$ or $\mathcal{G}$] have disappeared. 

\subsection{Some comments about the proper-time series expansion}
The asymptotic expansion (proper-time series expansion) that we are using here [see Eqs. \eqref{QED-expansion} and \eqref{expansion - 1}] is an expansion that group terms with the same mass dimension (remember that the coefficients $O_n$ and $a_n$ have mass dimension $2n$). This expansion is closely related with the adiabatic expansion of the scalar field modes, widely used in cosmology [see Ref. \cite{rio15} for the equivalence in curved spacetime, and \cite{FNP20} for the adiabatic counterpart of the $R$-summed expansion given in \eqref{R-summed exp}]. In its normal form, it is commonly used to renormalize physical quantities. It is important to recall that this expansion is not a derivative expansion (i.e., an expansion that groups terms with the same number of derivatives; see, for example, Ref. \cite{igor}). %The factorization proposed in Eq. \eqref{relation - 1} does not convert the proper-time expansion into a derivative expansion.  
For a detailed explanation of the differences between both expansions see %Sections 1.2.4, 1.3, 2.3 and 2.4 in 
Ref. \cite{Dunne1}.

\vspace{-0.2cm}
\section{Main consequences} \label{consequences}

Our factorization, allow us to predict or reproduce some interesting physical consequences. 
First, it is possible to find some exactly solvable electromagnetic backgrounds. For electric and magnetic fields pointing in the $\hat z $ direction that depend arbitrarily on the  light-cone coordinate $x^+=(t+z)$, we find that $\bar O_0=1$ and $\bar O_{n>0}=0$. Therefore, the  exact form of the unrenormalized effective Lagrangian should be (in agreement with Refs. \cite{Woodard1,Woodard2,Ilderton})
  \be     \mathcal {L}_{scalar}^{(1)}= -\frac{1}{16\pi^2} \int_0^\infty \frac{ds}{s^3} e^{-im^2 s}\frac{e^2s^2E(x^+)B(x^+)}{\sinh esE(x^+) \sin esB(x^+)}.   
 \label{Lscalar3} \ee 
For a single electromagnetic plane wave, $\mathcal{F}(x)=0$ and $\mathcal{G}(x)=0$, therefore,  according to our conjecture, the one-loop effective action trivially vanishes, in agreement with the result obtained in Ref. \cite{Schwinger} by other methods.
%which means that there are not quantum corrections to the classical Lagrangian, as explicitly proved in Ref. \cite{Schwinger}.
Second, our results seem to be consistent in presence of  gravity. \footnote{%We note that 
For $\nabla_\rho F^{\mu \nu}=0$ the factorization of the Euler-Heisenberg Lagrangian was found in Refs. \cite{Avramidi-Fucci1,Avramidi-Fucci2}.}. 
In this case, it is possible to perform a double factorization (the exponential $R(x)$ factorization is ensured\cite{Jack-Parker}),
 \be
 g(x,is)=e^{-is (\xi -\tfrac{1}{6})R(x)} {\small \Bigg[\det \left( \frac{esF(x)}{\sinh (esF(x))}\right )\Bigg]^{1/2}} \tilde g(x;is). \ee
 Finally, the conjecture allows us to make some general predictions regarding  the Schwinger's formula for the pair production rate.  The factorization suggests that the poles of the imaginary part of the one-loop effective Lagrangian are located at the same points as in the  constant electric field case $\tau_n=n\pi/|eE(x)|$.
 \vspace{-0.25cm}
\section{Summary and conclusions} \label{summary}

The one-loop (scalar) QED effective-Lagrangian is, in general, unknown. However, it is possible to obtain an  adiabatic expansion of it in terms  of the functions $g(x;is)$ defined in (\ref{gs}).  We have proposed an  alternative adiabatic expansion $\bar g(x;is)$, which encapsulates in a global pre-factor all its dependence on the electromagnetic invariants $\mathcal F(x)$ and $\mathcal G(x)$.  The form of the non-perturbative factor involved in this partial resummation is just the Heisenberg-Euler Lagrangian for QED, but with the electric and magnetic fields depending arbitrarily on spacetime coordinates. The new expansion does not contain terms that vanish when $\mathcal F(x)$ and $\mathcal G(x)$ are replaced by zero. This factorization allows us to obtain some  exact solutions and seems to be consistent in presence of gravity. In this contribution, we have focused on the scalar case, but our conjecture is also valid for the spin-$\frac{1}{2}$ case. We expect our results to be consistent for non-abelian gauge backgrounds. Most of the computations in this paper have been done with the help of the xAct package of the {\it Mathematica Software}\cite{programs}. 
This work was supported in part by Spanish Ministerio de Economia, Industria y Competitividad Grants No. FIS2017-84440-C2-1-P(MINECO/FEDER, EU), No. FIS2017-91161-EXP and
the project PROMETEO/2020/079 (Generalitat Valenciana). S. P. is supported by a Ph.D. fellowship, Grant No. FPU16/05287.

\appendix{} \label{apA}

 Here we give the coefficients $O_n$ of the original  expansion  $g(x;is)$.  $O_0=1,$  \bea 
O_{1}&=&0, \,\,\,  O_{2}=- {\bf \tfrac{e^2}{6} F_{\kappa \lambda } F^{\kappa \lambda }}, \,\,\,O_{3}=- \tfrac{e^2}{20} \partial_{\mu }F_{\kappa \lambda } \partial^{\mu }F^{\kappa \lambda },\\
\nonumber\\
O_{4}&=&{\bf \tfrac{e^4}{15} F_{\kappa }{}^{\mu } F^{\kappa \lambda } F_{\lambda }{}^{\nu } F_{\mu \nu }} + \bf{\tfrac{e^4}{12} F_{\kappa \lambda } F^{\kappa \lambda } F_{\mu \nu } F^{\mu \nu }} \nonumber\\&\,&- \tfrac{e^2}{70} \partial_{\nu }\partial_{\mu }F_{\kappa \lambda } \partial^{\nu }\partial^{\mu }F^{\kappa \lambda },\\
 \nonumber\\
O_5&=&\tfrac{2e^4}{7} F^{\kappa \lambda } F^{\mu \nu } \partial_{\lambda }F_{\nu \rho } \partial_{\mu }F_{\kappa }{}^{\rho } -  \tfrac{4e^4}{63} F_{\kappa }{}^{\mu } F^{\kappa \lambda } \partial_{\lambda }F^{\nu \rho } \partial_{\mu }F_{\nu \rho } \nonumber \\
&&-  \tfrac{e^4}{9} F_{\kappa }{}^{\mu } F^{\kappa \lambda } F^{\nu \rho } \partial_{\mu }\partial_{\lambda }F_{\nu \rho } -  \tfrac{16e^4}{63} F^{\kappa \lambda } F^{\mu \nu } \partial_{\mu }F_{\kappa }{}^{\rho } \partial_{\nu }F_{\lambda \rho } \nonumber\\
&&+ \tfrac{5e^4}{18} F^{\kappa \lambda } F^{\mu \nu } \partial_{\rho }F_{\mu \nu } \partial^{\rho }F_{\kappa \lambda } + \tfrac{34e^4}{189} F^{\kappa \lambda } F^{\mu \nu } \partial_{\nu }F_{\lambda \rho } \partial^{\rho }F_{\kappa \mu } \nonumber\\
&&+ \tfrac{25e^4}{189} F^{\kappa \lambda } F^{\mu \nu } \partial_{\rho }F_{\lambda \nu } \partial^{\rho }F_{\kappa \mu } + \tfrac{4e^4}{21} F_{\kappa }{}^{\mu } F^{\kappa \lambda } \partial_{\rho }F_{\mu \nu } \partial^{\rho }F_{\lambda }{}^{\nu } \nonumber\\
&&+ {\bf \tfrac{e^4}{12} F_{\kappa \lambda } F^{\kappa \lambda } \partial_{\rho }F_{\mu \nu } \partial^{\rho }F^{\mu \nu } }-  \tfrac{e^2}{252} \partial_{\rho }\partial_{\nu }\partial_{\mu }F_{\kappa \lambda } \partial^{\rho }\partial^{\nu }\partial^{\mu }F^{\kappa \lambda },\\
\nonumber\\
O_6&=& -{\bf \tfrac{8}{63} F_{\kappa }{}^{\mu } F^{\kappa \lambda } F_{\lambda }{}^{\nu } F_{\mu }{}^{\rho } F_{\nu }{}^{\sigma } F_{\rho \sigma }} -  {\bf \tfrac{1}{6} F_{\kappa \lambda } F^{\kappa \lambda } F_{\mu }{}^{\rho } F^{\mu \nu } F_{\nu }{}^{\sigma } F_{\rho \sigma }} \nonumber\\&&- {\bf\tfrac{5}{72} F_{\kappa \lambda } F^{\kappa \lambda } F_{\mu \nu } F^{\mu \nu } F_{\rho \sigma } F^{\rho \sigma }} + \tfrac{397}{3465} F^{\kappa \lambda } \partial_{\kappa }F^{\mu \nu } \partial_{\mu }F^{\rho \sigma } \partial_{\nu }\partial_{\lambda }F_{\rho \sigma } \nonumber\\&&+ \tfrac{1}{63} F^{\kappa \lambda } F^{\mu \nu } \partial_{\mu }\partial_{\kappa }F^{\rho \sigma } \partial_{\nu }\partial_{\lambda }F_{\rho \sigma } + \tfrac{187}{1260} \partial_{\mu }F^{\rho \sigma } \partial^{\mu }F^{\kappa \lambda } \partial_{\nu }F_{\rho \sigma } \partial^{\nu }F_{\kappa \lambda }\nonumber\\&& -  \tfrac{1079}{3465} F^{\kappa \lambda } \partial_{\mu }\partial_{\lambda }F_{\rho \sigma } \partial_{\nu }F^{\rho \sigma } \partial^{\nu }F_{\kappa }{}^{\mu } -  \tfrac{43}{3465} F^{\kappa \lambda } \partial_{\mu }F^{\rho \sigma } \partial_{\nu }\partial_{\lambda }F_{\rho \sigma } \partial^{\nu }F_{\kappa }{}^{\mu }\nonumber\\&& -  \tfrac{4}{35} F^{\kappa \lambda } \partial_{\lambda }F^{\rho \sigma } \partial_{\nu }\partial_{\mu }F_{\rho \sigma } \partial^{\nu }F_{\kappa }{}^{\mu } -  \tfrac{101}{3465} \partial_{\lambda }F_{\rho \sigma } \partial^{\mu }F^{\kappa \lambda } \partial_{\nu }F_{\mu }{}^{\sigma } \partial^{\rho }F_{\kappa }{}^{\nu } \nonumber\\&&+ \tfrac{43}{3465} \partial_{\mu }F_{\nu \sigma } \partial^{\mu }F^{\kappa \lambda } \partial_{\rho }F_{\lambda }{}^{\sigma } \partial^{\rho }F_{\kappa }{}^{\nu } -  \tfrac{2}{21} F^{\kappa \lambda } \partial_{\rho }F_{\mu }{}^{\sigma } \partial^{\rho }F^{\mu \nu } \partial_{\sigma }\partial_{\nu }F_{\kappa \lambda } \nonumber\\&&-  \tfrac{4}{21} F^{\kappa \lambda } F^{\mu \nu } \partial_{\sigma }\partial_{\rho }\partial_{\lambda }F_{\mu \nu } \partial^{\sigma }F_{\kappa }{}^{\rho } + \tfrac{1486}{3465} \partial_{\mu }F_{\nu \sigma } \partial^{\mu }F^{\kappa \lambda } \partial^{\rho }F_{\kappa }{}^{\nu } \partial^{\sigma }F_{\lambda \rho } \nonumber\\&&+ \tfrac{428}{1155} \partial^{\mu }F^{\kappa \lambda } \partial^{\rho }F_{\kappa }{}^{\nu } \partial_{\sigma }F_{\mu \nu } \partial^{\sigma }F_{\lambda \rho } + \tfrac{3}{35} \partial_{\mu }F_{\kappa }{}^{\nu } \partial^{\mu }F^{\kappa \lambda } \partial_{\sigma }F_{\nu \rho } \partial^{\sigma }F_{\lambda }{}^{\rho } \nonumber\\&&+ \tfrac{6}{35} F^{\kappa \lambda } \partial^{\nu }F_{\kappa }{}^{\mu } \partial_{\sigma }\partial_{\nu }F_{\mu \rho } \partial^{\sigma }F_{\lambda }{}^{\rho } + \tfrac{13}{42} F^{\kappa \lambda } \partial^{\rho }F^{\mu \nu } \partial_{\sigma }\partial_{\rho }F_{\kappa \lambda } \partial^{\sigma }F_{\mu \nu } \nonumber\\&&-  \tfrac{2}{35} \partial^{\mu }F^{\kappa \lambda } \partial^{\nu }F_{\kappa \lambda } \partial_{\sigma }F_{\nu \rho } \partial^{\sigma }F_{\mu }{}^{\rho } + \tfrac{64}{231} F^{\kappa \lambda } \partial_{\kappa }F^{\mu \nu } \partial_{\sigma }\partial_{\lambda }F_{\nu \rho } \partial^{\sigma }F_{\mu }{}^{\rho } \nonumber\\&&+ \tfrac{118}{693} F^{\kappa \lambda } \partial^{\nu }F_{\kappa }{}^{\mu } \partial_{\sigma }\partial_{\lambda }F_{\nu \rho } \partial^{\sigma }F_{\mu }{}^{\rho } -  \tfrac{302}{3465} F^{\kappa \lambda } \partial_{\kappa }F^{\mu \nu } \partial_{\sigma }\partial_{\nu }F_{\lambda \rho } \partial^{\sigma }F_{\mu }{}^{\rho } \nonumber\\&&+ \tfrac{1388}{3465} F^{\kappa \lambda } \partial^{\nu }F_{\kappa }{}^{\mu } \partial_{\sigma }\partial_{\nu }F_{\lambda \rho } \partial^{\sigma }F_{\mu }{}^{\rho } + \tfrac{344}{1155} F^{\kappa \lambda } \partial_{\kappa }F^{\mu \nu } \partial_{\sigma }\partial_{\rho }F_{\lambda \nu } \partial^{\sigma }F_{\mu }{}^{\rho } \nonumber\\&&+ \tfrac{398}{3465} F^{\kappa \lambda } \partial^{\nu }F_{\kappa }{}^{\mu } \partial_{\sigma }\partial_{\rho }F_{\lambda \nu } \partial^{\sigma }F_{\mu }{}^{\rho } -  \tfrac{62}{165} F^{\kappa \lambda } \partial^{\nu }F_{\kappa }{}^{\mu } \partial_{\sigma }\partial_{\lambda }F_{\mu \rho } \partial^{\sigma }F_{\nu }{}^{\rho } \nonumber\\&&+ \tfrac{76}{693} F^{\kappa \lambda } \partial^{\nu }F_{\kappa }{}^{\mu } \partial_{\sigma }\partial_{\mu }F_{\lambda \rho } \partial^{\sigma }F_{\nu }{}^{\rho } -  \tfrac{326}{1155} F^{\kappa \lambda } \partial^{\nu }F_{\kappa }{}^{\mu } \partial_{\sigma }\partial_{\rho }F_{\lambda \mu } \partial^{\sigma }F_{\nu }{}^{\rho } \nonumber\\&&+ \tfrac{1}{40} \partial_{\mu }F_{\kappa \lambda } \partial^{\mu }F^{\kappa \lambda } \partial_{\sigma }F_{\nu \rho } \partial^{\sigma }F^{\nu \rho } + \tfrac{1}{3} F^{\kappa \lambda } \partial^{\mu }F_{\kappa \lambda } \partial_{\sigma }\partial_{\mu }F_{\nu \rho } \partial^{\sigma }F^{\nu \rho } \nonumber\\&&-  \tfrac{1}{15} F_{\kappa }{}^{\mu } F^{\kappa \lambda } \partial_{\sigma }\partial_{\mu }\partial_{\lambda }F_{\nu \rho } \partial^{\sigma }F^{\nu \rho } -  \tfrac{2}{105} F_{\kappa }{}^{\mu } F^{\kappa \lambda } \partial_{\sigma }\partial_{\mu }F_{\nu \rho } \partial^{\sigma }\partial_{\lambda }F^{\nu \rho } \nonumber\\&&+ \tfrac{43}{231} F^{\kappa \lambda } F^{\mu \nu } \partial_{\sigma }\partial_{\lambda }F_{\nu \rho } \partial^{\sigma }\partial_{\mu }F_{\kappa }{}^{\rho } -  \tfrac{43}{231} F^{\kappa \lambda } F^{\mu \nu } \partial_{\sigma }\partial_{\nu }F_{\lambda \rho } \partial^{\sigma }\partial_{\mu }F_{\kappa }{}^{\rho } \nonumber\\&&+ \tfrac{1}{6} F^{\kappa \lambda } F^{\mu \nu } \partial_{\sigma }\partial_{\rho }F_{\mu \nu } \partial^{\sigma }\partial^{\rho }F_{\kappa \lambda } + \tfrac{103}{693} F^{\kappa \lambda } F^{\mu \nu } \partial_{\sigma }\partial_{\nu }F_{\lambda \rho } \partial^{\sigma }\partial^{\rho }F_{\kappa \mu } \nonumber\\&&+ \tfrac{46}{693} F^{\kappa \lambda } F^{\mu \nu } \partial_{\sigma }\partial_{\rho }F_{\lambda \nu } \partial^{\sigma }\partial^{\rho }F_{\kappa \mu } + \tfrac{2}{21} F_{\kappa }{}^{\mu } F^{\kappa \lambda } \partial_{\sigma }\partial_{\rho }F_{\mu \nu } \partial^{\sigma }\partial^{\rho }F_{\lambda }{}^{\nu } \nonumber\\&&+ {\bf \tfrac{1}{28} F_{\kappa \lambda } F^{\kappa \lambda } \partial_{\sigma }\partial_{\rho }F_{\mu \nu } \partial^{\sigma }\partial^{\rho }F^{\mu \nu }} -  \tfrac{1}{924} \partial_{\sigma }\partial_{\rho }\partial_{\nu }\partial_{\mu }F_{\kappa \lambda } \partial^{\sigma }\partial^{\rho }\partial^{\nu }\partial^{\mu }F^{\kappa \lambda }.
\eea
The terms that have to disappear when computing the modified expansion $\bar g(x;is)$ for our conjecture to be satisfied appeared highlighted.

\appendix{} \label{apB}
Here we give the coefficients $\bar O_n$ of the  $(\mathcal{F},\mathcal{G})-$summed expansion $\bar g(x;is)$, $\bar O_0=1,$

 \bea
\bar O_{1}&=&0, \,\,\,  \bar O_{2}=0, \,\,\,\bar O_{3}=- \tfrac{e^2}{20} \partial_{\mu }F_{\kappa \lambda } \partial^{\mu }F^{\kappa \lambda },\\
\nonumber\\
\bar O_{4}&=&- \tfrac{e^2}{70} \partial_{\nu }\partial_{\mu }F_{\kappa \lambda } \partial^{\nu }\partial^{\mu }F^{\kappa \lambda },\\
\nonumber\\
\bar O_5&=&\tfrac{2e^4}{7} F^{\kappa \lambda } F^{\mu \nu } \partial_{\lambda }F_{\nu \rho } \partial_{\mu }F_{\kappa }{}^{\rho } -  \tfrac{4e^4}{63} F_{\kappa }{}^{\mu } F^{\kappa \lambda } \partial_{\lambda }F^{\nu \rho } \partial_{\mu }F_{\nu \rho } \nonumber \\
&&-  \tfrac{e^4}{9} F_{\kappa }{}^{\mu } F^{\kappa \lambda } F^{\nu \rho } \partial_{\mu }\partial_{\lambda }F_{\nu \rho } -  \tfrac{16e^4}{63} F^{\kappa \lambda } F^{\mu \nu } \partial_{\mu }F_{\kappa }{}^{\rho } \partial_{\nu }F_{\lambda \rho } \nonumber\\
&&+ \tfrac{5e^4}{18} F^{\kappa \lambda } F^{\mu \nu } \partial_{\rho }F_{\mu \nu } \partial^{\rho }F_{\kappa \lambda } + \tfrac{34e^4}{189} F^{\kappa \lambda } F^{\mu \nu } \partial_{\nu }F_{\lambda \rho } \partial^{\rho }F_{\kappa \mu } \nonumber\\
&&+ \tfrac{25e^4}{189} F^{\kappa \lambda } F^{\mu \nu } \partial_{\rho }F_{\lambda \nu } \partial^{\rho }F_{\kappa \mu } + \tfrac{4e^4}{21} F_{\kappa }{}^{\mu } F^{\kappa \lambda } \partial_{\rho }F_{\mu \nu } \partial^{\rho }F_{\lambda }{}^{\nu } \nonumber\\
&& -  \tfrac{e^2}{252} \partial_{\rho }\partial_{\nu }\partial_{\mu }F_{\kappa \lambda } \partial^{\rho }\partial^{\nu }\partial^{\mu }F^{\kappa \lambda },
\\ \nonumber\\
\bar O_6&=&  + \tfrac{397}{3465} F^{\kappa \lambda } \partial_{\kappa }F^{\mu \nu } \partial_{\mu }F^{\rho \sigma } \partial_{\nu }\partial_{\lambda }F_{\rho \sigma } + \tfrac{1}{63} F^{\kappa \lambda } F^{\mu \nu } \partial_{\mu }\partial_{\kappa }F^{\rho \sigma } \partial_{\nu }\partial_{\lambda }F_{\rho \sigma } \nonumber\\&&+ \tfrac{187}{1260} \partial_{\mu }F^{\rho \sigma } \partial^{\mu }F^{\kappa \lambda } \partial_{\nu }F_{\rho \sigma } \partial^{\nu }F_{\kappa \lambda }-  \tfrac{1079}{3465} F^{\kappa \lambda } \partial_{\mu }\partial_{\lambda }F_{\rho \sigma } \partial_{\nu }F^{\rho \sigma } \partial^{\nu }F_{\kappa }{}^{\mu } \nonumber\\&&-  \tfrac{43}{3465} F^{\kappa \lambda } \partial_{\mu }F^{\rho \sigma } \partial_{\nu }\partial_{\lambda }F_{\rho \sigma } \partial^{\nu }F_{\kappa }{}^{\mu } -  \tfrac{4}{35} F^{\kappa \lambda } \partial_{\lambda }F^{\rho \sigma } \partial_{\nu }\partial_{\mu }F_{\rho \sigma } \partial^{\nu }F_{\kappa }{}^{\mu } \nonumber\\&&-  \tfrac{101}{3465} \partial_{\lambda }F_{\rho \sigma } \partial^{\mu }F^{\kappa \lambda } \partial_{\nu }F_{\mu }{}^{\sigma } \partial^{\rho }F_{\kappa }{}^{\nu } + \tfrac{43}{3465} \partial_{\mu }F_{\nu \sigma } \partial^{\mu }F^{\kappa \lambda } \partial_{\rho }F_{\lambda }{}^{\sigma } \partial^{\rho }F_{\kappa }{}^{\nu } \nonumber\\&&-  \tfrac{2}{21} F^{\kappa \lambda } \partial_{\rho }F_{\mu }{}^{\sigma } \partial^{\rho }F^{\mu \nu } \partial_{\sigma }\partial_{\nu }F_{\kappa \lambda } \-  \tfrac{4}{21} F^{\kappa \lambda } F^{\mu \nu } \partial_{\sigma }\partial_{\rho }\partial_{\lambda }F_{\mu \nu } \partial^{\sigma }F_{\kappa }{}^{\rho } \nonumber\\&&+ \tfrac{1486}{3465} \partial_{\mu }F_{\nu \sigma } \partial^{\mu }F^{\kappa \lambda } \partial^{\rho }F_{\kappa }{}^{\nu } \partial^{\sigma }F_{\lambda \rho } + \tfrac{428}{1155} \partial^{\mu }F^{\kappa \lambda } \partial^{\rho }F_{\kappa }{}^{\nu } \partial_{\sigma }F_{\mu \nu } \partial^{\sigma }F_{\lambda \rho } \nonumber\\&&+ \tfrac{3}{35} \partial_{\mu }F_{\kappa }{}^{\nu } \partial^{\mu }F^{\kappa \lambda } \partial_{\sigma }F_{\nu \rho } \partial^{\sigma }F_{\lambda }{}^{\rho } + \tfrac{6}{35} F^{\kappa \lambda } \partial^{\nu }F_{\kappa }{}^{\mu } \partial_{\sigma }\partial_{\nu }F_{\mu \rho } \partial^{\sigma }F_{\lambda }{}^{\rho } \nonumber\\&&+ \tfrac{13}{42} F^{\kappa \lambda } \partial^{\rho }F^{\mu \nu } \partial_{\sigma }\partial_{\rho }F_{\kappa \lambda } \partial^{\sigma }F_{\mu \nu } -  \tfrac{2}{35} \partial^{\mu }F^{\kappa \lambda } \partial^{\nu }F_{\kappa \lambda } \partial_{\sigma }F_{\nu \rho } \partial^{\sigma }F_{\mu }{}^{\rho } \nonumber\\&&+ \tfrac{64}{231} F^{\kappa \lambda } \partial_{\kappa }F^{\mu \nu } \partial_{\sigma }\partial_{\lambda }F_{\nu \rho } \partial^{\sigma }F_{\mu }{}^{\rho } + \tfrac{118}{693} F^{\kappa \lambda } \partial^{\nu }F_{\kappa }{}^{\mu } \partial_{\sigma }\partial_{\lambda }F_{\nu \rho } \partial^{\sigma }F_{\mu }{}^{\rho } \nonumber\\&& -  \tfrac{302}{3465} F^{\kappa \lambda } \partial_{\kappa }F^{\mu \nu } \partial_{\sigma }\partial_{\nu }F_{\lambda \rho } \partial^{\sigma }F_{\mu }{}^{\rho } + \tfrac{1388}{3465} F^{\kappa \lambda } \partial^{\nu }F_{\kappa }{}^{\mu } \partial_{\sigma }\partial_{\nu }F_{\lambda \rho } \partial^{\sigma }F_{\mu }{}^{\rho } \nonumber\\&& + \tfrac{344}{1155} F^{\kappa \lambda } \partial_{\kappa }F^{\mu \nu } \partial_{\sigma }\partial_{\rho }F_{\lambda \nu } \partial^{\sigma }F_{\mu }{}^{\rho } + \tfrac{398}{3465} F^{\kappa \lambda } \partial^{\nu }F_{\kappa }{}^{\mu } \partial_{\sigma }\partial_{\rho }F_{\lambda \nu } \partial^{\sigma }F_{\mu }{}^{\rho } \nonumber\\&& -  \tfrac{62}{165} F^{\kappa \lambda } \partial^{\nu }F_{\kappa }{}^{\mu } \partial_{\sigma }\partial_{\lambda }F_{\mu \rho } \partial^{\sigma }F_{\nu }{}^{\rho } + \tfrac{76}{693} F^{\kappa \lambda } \partial^{\nu }F_{\kappa }{}^{\mu } \partial_{\sigma }\partial_{\mu }F_{\lambda \rho } \partial^{\sigma }F_{\nu }{}^{\rho } \nonumber\\&&-  \tfrac{326}{1155} F^{\kappa \lambda } \partial^{\nu }F_{\kappa }{}^{\mu } \partial_{\sigma }\partial_{\rho }F_{\lambda \mu } \partial^{\sigma }F_{\nu }{}^{\rho } + \tfrac{1}{40} \partial_{\mu }F_{\kappa \lambda } \partial^{\mu }F^{\kappa \lambda } \partial_{\sigma }F_{\nu \rho } \partial^{\sigma }F^{\nu \rho } \nonumber\\&&+ \tfrac{1}{3} F^{\kappa \lambda } \partial^{\mu }F_{\kappa \lambda } \partial_{\sigma }\partial_{\mu }F_{\nu \rho } \partial^{\sigma }F^{\nu \rho } -  \tfrac{1}{15} F_{\kappa }{}^{\mu } F^{\kappa \lambda } \partial_{\sigma }\partial_{\mu }\partial_{\lambda }F_{\nu \rho } \partial^{\sigma }F^{\nu \rho } \nonumber\\&&-  \tfrac{2}{105} F_{\kappa }{}^{\mu } F^{\kappa \lambda } \partial_{\sigma }\partial_{\mu }F_{\nu \rho } \partial^{\sigma }\partial_{\lambda }F^{\nu \rho }+ \tfrac{43}{231} F^{\kappa \lambda } F^{\mu \nu } \partial_{\sigma }\partial_{\lambda }F_{\nu \rho } \partial^{\sigma }\partial_{\mu }F_{\kappa }{}^{\rho } \nonumber\\&&-  \tfrac{43}{231} F^{\kappa \lambda } F^{\mu \nu } \partial_{\sigma }\partial_{\nu }F_{\lambda \rho } \partial^{\sigma }\partial_{\mu }F_{\kappa }{}^{\rho } + \tfrac{1}{6} F^{\kappa \lambda } F^{\mu \nu } \partial_{\sigma }\partial_{\rho }F_{\mu \nu } \partial^{\sigma }\partial^{\rho }F_{\kappa \lambda } \nonumber\\&&+ \tfrac{103}{693} F^{\kappa \lambda } F^{\mu \nu } \partial_{\sigma }\partial_{\nu }F_{\lambda \rho } \partial^{\sigma }\partial^{\rho }F_{\kappa \mu } + \tfrac{46}{693} F^{\kappa \lambda } F^{\mu \nu } \partial_{\sigma }\partial_{\rho }F_{\lambda \nu } \partial^{\sigma }\partial^{\rho }F_{\kappa \mu } \nonumber\\&&+ \tfrac{2}{21} F_{\kappa }{}^{\mu } F^{\kappa \lambda } \partial_{\sigma }\partial_{\rho }F_{\mu \nu } \partial^{\sigma }\partial^{\rho }F_{\lambda }{}^{\nu } -  \tfrac{1}{924} \partial_{\sigma }\partial_{\rho }\partial_{\nu }\partial_{\mu }F_{\kappa \lambda } \partial^{\sigma }\partial^{\rho }\partial^{\nu }\partial^{\mu }F^{\kappa \lambda }.
\eea

%\bibliographystyle{ws-procs961x669}
%\bibliography{ws-pro-sample}

%Non BiBTeX users can list down their references as:

\end{document}